\documentclass[prb,twocolumn,showpacs,amsmath,superscriptaddress,amssymb]{revtex4}

\usepackage{graphicx}
\usepackage{dcolumn}
\usepackage{bm}

\begin{document}


\title{Landau-Zener tunneling problem for Bloch states
}
\author{Ryuji Takahashi}
\email{ryuji.takahashi@riken.jp}

\affiliation{Condensed Matter Theory Laboratory, RIKEN, Wako, Saitama 351-0198, Japan}
\author{Naoyuki Sugimoto}
\affiliation{Department of Applied Physics, The University of Tokyo
}
\date{\today}

\begin{abstract}
We study the Landau-Zener tunneling problem for particles bound in periodic lattice insulators. To this end, we construct the path integral based on the Bloch and Wannier functions in the presence with an external force, and the transition amplitude is calculated for the Su-Schrieffer-Heeger model. We find that the tunneling probability in bulk periodic systems becomes drastically larger than that by the Landau-Zener formula. This enhancement is prominent for small values of the external field or small hopping integral compared with the gap, and comes from the difference between the Dirac and the periodic dispersions. In addition, when the lattice effect is strong, another analytical formula of the tunneling probability is given with a different behavior from the Landau-Zener formula. Finally, we discuss the observation scheme for the lattice effect.
\end{abstract}

\maketitle

\section{Introduction}
%

The tunneling phenomenon represents one of the most fundamental features of quantum mechanics.
 The transition of the particle between the potential barrier cannot be described by classical physics.
The Landau-Zener (LZ) problem gives a simple description for the non-adiabatic transition between two levels~\cite{Landau32,LandauB, Zener32} . Two-state systems are ubiquitously obtained, and this problem has been studied in various quantum systems, e.g., optical systems~\cite{Spreeuw90, Bouwmeester95}, bosonic systems~\cite{ NiuQ00, Liu02,Chen11}, chemical bonds~\cite{ChildB, NikitinB}, and Zener breakdown with metal-insulator transitions of semiconductors~\cite{ZimanB72, Asamitsu97, Miyano97,Yamanouchi99,Taguchi00, Sugimoto08,OkaT05}.

The tunneling probability of the LZ problem is given by the celebrated LZ formula, which is derived by the Wentzel-Kramers-Brillouin (WKB) method  in a conventional way\cite{SakuraiB}.
Typically, for the electron in solids, this widely-used formula is derived by the low-energy model based on the $k\cdot p$ approximation. 
The low-energy Hamiltonian is introduced as $H_{\mathrm{D}}=vk\alpha_{z} +\Delta\alpha_{x}$, with the momentum $k$, the velocity $v$, and the gap $\Delta$.
$\boldsymbol{\alpha}$ are the Pauli matrices for the band, and the chemical potential is zero.
 The energy dispersion is given as $\xi_{k,\pm}=\pm \xi _{k}$ with $\xi _{k}=\sqrt{\Delta^2+v^2k^2}$.
 In the semiclassical picture of the system with the external force $V(x)=-Fx$, the wavenumber depends on time $k= Ft+$ const., and the kinetic energy of the Dirac Hamiltonian $vk\alpha_{z}$ can be regarded as the time-dependent level separation of the LZ problem.
 Therefore, this one-dimensional (1D) massive Dirac model is true to the system originally given by Zener.
In the transition process, $k$ becomes imaginary, and the energy changes from $-\Delta$ to $\Delta$ by the force $F$~\cite{ZimanB72} with the spatial transition from $-X_{0}$ to $X_{0}\left(=\frac{\Delta}{F}\right)$ by the total energy conservation law.
 The probability by the LZ formula is given as 
 \begin{eqnarray}
P_{\mathrm{D}}= \mathrm{e}^{2i S_{\mathrm{D}}},
\label{eq:PLZ}
\end{eqnarray}
with $i S_{\mathrm{D}}=-\frac{\pi\Delta^2}{2Fv}$.

On the other hand, especially for bound particles, we cannot always obtain the linear separation. 
In lattice systems, the level separation for the LZ problem is given by gapless terms of the Hamiltonian, and the dispersion has a nonlinear form by reflecting the spatial periodicity of the system. Namely, it is not sufficient to describe features peculiar to the particles in lattice systems by the LZ formula for the linear approximation.
Moreover, approximation methods should be carefully used for calculations of the tunneling amplitude, since the transition occurs between distant energy levels, and therefore non-perturbative methods are favorable for the description of tunneling phenomena.

In this paper,
  we consider the LZ problem for periodic lattice systems by the external force in real space. We predict that the tunneling probability in bulk becomes drastically larger than that by the LZ formula and show that the lattice effect on the LZ problem cannot be neglected for the small velocity or small force. 
 To this end, 
 we construct the path integral~\cite{FeynmanB} of the Bloch states, and the transition probability by the semiclassical contribution is analytically calculated by using the instanton method~\cite{ColemanB} for the Su-Schrieffer-Heeger (SSH) lattice model \cite{Su79, Su80}.
 Our findings imply the higher-order effect plays a large role in various tunneling phenomena.


\begin{figure}[htbp]
 \begin{center}
 \includegraphics[width=80mm]{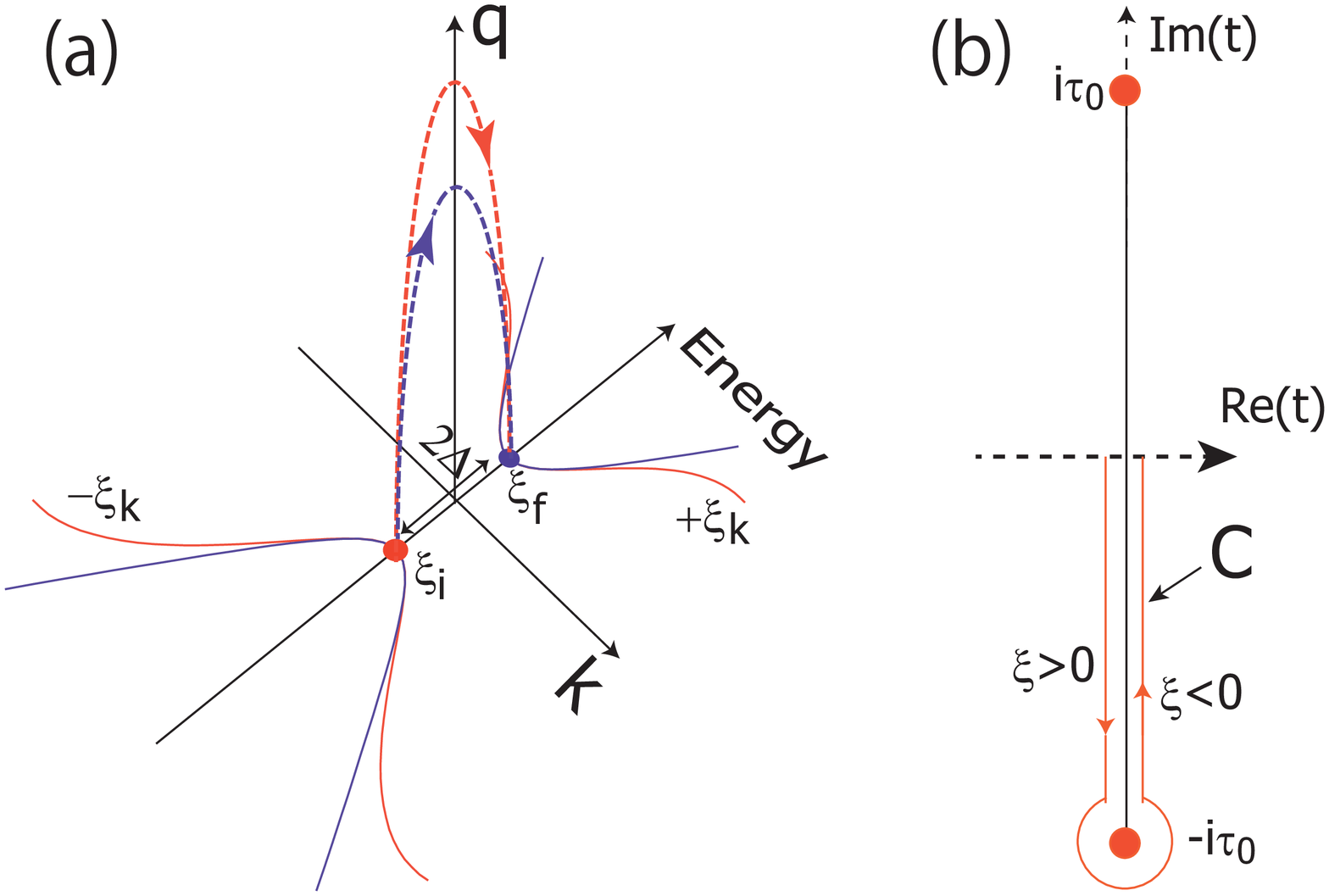}
 \caption{ (a) Schematic of the energy dispersions and non-adiabatic transition process.
The energy dispersions are shown on the $k$ axis, for the Dirac Hamiltonian (solid blue curve) and tight-binding Hamiltonian (\ref{eq:Ham1}) (solid red curve), respectively.
Their transition processes are shown in the dotted curves on at the imaginary wavenumber $k\to iq$.
  The initial and final energies are on the lower and upper band edges:  $\xi_{i}=-\xi_{k=0}=-\Delta$ and $\xi_{f}=+\xi_{k=0}=\Delta$.
 Due to the higher-order effect, there is a difference in the tunneling path. 
 (b) Schematic of the time contour of the instanton.
 The two bold dots show the branching points $\pm \tau_{0}=\pm \frac{1}{F}\mathrm{Arcsinh}\frac{\Delta}{v}$ with $\xi_{F\tau_{0}}=0$, and the solid line between the two dots is the branch cut.
 Along the contour, $\xi$ changes its sign by the transition to another sheet of the Riemann surface.
}
\label{fig:system1D}
\end{center}\end{figure}




\section{Path integral of the Bloch states in one dimension}
The path integral is a well-established method for the calculation of the transition amplitude of the quantum states. 
Here, we construct the calculation of the amplitude by using the Bloch and Wannier states in 1D systems, and we obtain the path integral described by the lattice coordinate $X$, wavenumber $k$, and the spinor $\eta$ representing the degrees of freedom of the system.
To this end, 
we first assume that the quantum particle in the lattice is described by the Bloch states $|\psi(n,k)\rangle$ with the band index  $n=1,2,\dots,M$ (the number of bands: $M$).
Without the external field, the Hamiltonian $H_k$ maintains periodicity, and its eigenstates take the Bloch form,
 $H_k|\psi(n,k)\rangle =\xi_{nk}|\psi(n,k)\rangle$ with the band energy $\xi_{nk}$.
In addition, we employ the the Wannier states:
  \begin{eqnarray}
 |n,X \rangle=\frac{1}{\sqrt{N_{a}}}\sum_{k,n'}
 \mathrm{e}^{-ik X}\mathcal{U}_{n'nk}|\psi(n',k)\rangle,
\label{eq:wf}
\end{eqnarray}
with the number of atoms $N_{a}$. 
$\mathcal{U}_{k}$ is the unitary matrix defined as
 \begin{eqnarray}
\mathcal{U}_{k} = \mathcal{T}\exp\left(i\int^{k}\mathrm{d}k'\mathcal{A}_{k'}\right),
\end{eqnarray}
 where the connection is $\mathcal{A}_{n'n} =\langle u_{n'k} |i\partial_{k} u_{nk}\rangle$, and $|\psi(n,k)\rangle=\mathrm{e}^{ikx}|u_{n,k}\rangle$.
$\mathcal{T}$ is the time-order operator of the path integral.
Then the matrix element of the position operator $x$ is expressed by the lattice coordinate: $ \langle n', X|x|n,X \rangle =X$.

According to the path integral procedure\cite{FeynmanB}, we consider the finite time transition amplitude from $t=t_{i}$ to $t_{f}$,
 and the time interval is divided into $N$ with $\epsilon=\frac{t_{f}-t_{i}}{N}$.
We express the state at $t_{j}=\epsilon j+t_{i}$ as 
$
|\boldsymbol{\eta}_{ t _{j}},k_{j} \rangle=\sum_{n_{j}}\eta_{j}^{n_{j}}| \psi(n_{j},k_{j})
\rangle
$ by the spinor $\boldsymbol{\eta}_{j}$. The spinor $\boldsymbol{\eta}_{j}$ satisfies $\boldsymbol{\eta}_{j}^{\mu\dagger} \cdot\boldsymbol{\eta}_{j}^{\nu} =\delta_{\mu\nu}$, $^\forall j\in [1,\dots,N] $
for $\mu,\nu \in[1,\dots,M]$.
The transition matrix element $\mathcal{K}(\Upsilon_{f},\Upsilon_{i})$ between the initial state $\Upsilon _{i}\equiv (\boldsymbol{\eta}_{ t _{i}}, k_{i})$ and the final state $\Upsilon_{f}\equiv (\boldsymbol{\eta}_{ t _{f}} , k_{f})$ as
 \begin{eqnarray}
\mathcal{K}(\Upsilon _{f},\Upsilon _{i})
&=&\langle \boldsymbol{\eta}_{ t _{f}} ,k_{f}|\mathrm{e}^{-i\int^{t_{f}}_{t_{i}}\Tilde{H} \mathrm{d}t} |\boldsymbol{\eta}_{ t _{i}},k_{i} \rangle
\nonumber\\
&=&\int \mathcal{D} \Upsilon  \prod_{j=1}^{N}U_{j}= \int\mathcal{D}\Upsilon\mathrm{e}^{i\int_{t_{i}}^{t_{f}} \mathrm{d}t\mathcal{L}  }\label{eq:pathinteg1},
\end{eqnarray}
 with $ \mathcal{D} \Upsilon = \mathcal{D} k\mathcal{D}X\mathcal{D}\eta^{\dagger}\mathcal{D}\eta$.
 The total Hamiltonian is given  by the Bloch Hamiltonian and  the external potential $V=-Fx$: $\Tilde{H} =H_{k}+V(x)$.
 The component of the path integral is given as
 \begin{eqnarray}
 U_{j}  &=&\langle \boldsymbol{\eta}_{ t _{j+1}},k_{j+1}|\mathrm{e}^{-i\epsilon \Tilde{H}}|\boldsymbol{\eta}_{ t _{j}},k_{j} \rangle
 \nonumber\\
&=&\frac{1}{N_{a}} \sum_{m_{j},X_{j}} \langle \boldsymbol{\eta}_{ t _{j+1}},k_{j+1}|m_{j}, X_{j} \rangle\langle m_{j}, X_{j} | \mathrm{e}^{-i\epsilon \Tilde{H}}|\boldsymbol{\eta}_{ t _{j}},k_{j} \rangle
 \nonumber\\
&=& \frac{1}{N_{a}} \sum_{m_{j},X_{j}}
\mathrm{e}^{i\epsilon \mathcal{L}_{j}},
\end{eqnarray}
By calculating the transition process, the Lagrangian $\mathcal{L}$ is expressed for both the lattice coordinate and the wavenumber,
 \begin{eqnarray}
 \mathcal{L}_{j}
=-X_{j}\cdot \dot{k}_{j} 
-i\dot{\boldsymbol{\eta}}_{j}^{\dagger} \cdot {\boldsymbol{\eta}}_{j}
+ \dot{k}\langle \mathcal{A}_{k}\rangle -\langle \hat{\xi}_{k}\rangle-V(X_{j}),\label{eq:Lagwp}
\end{eqnarray}
where $\hat{\xi}_{k}=\mathrm{diag}(\xi_{1k},\xi_{2k},\dots,\xi_{Mk})$, the bracket and dot are defined as $\langle \mathcal{O} \rangle\equiv \boldsymbol{\eta}^{\dagger} \mathcal{O} \boldsymbol{\eta}$, and $\dot{\mathcal{O}}_{j}\equiv \frac{\mathcal{O}_{j}-\mathcal{O}_{j-1}}{\epsilon}$.
 Similar Lagrangians can be seen in the wave-packet theory~\cite{Culcer05, Xiao10}.
Owing to Eq.~(\ref{eq:wf}),  the external field in the Lagrangian~(\ref{eq:Lagwp}) is expressed by the lattice coordinate $V(x)\to V(X)=-FX$, and
then 
the path integral is expressed by the effective Lagrangian $\Tilde{\mathcal{L}}$ derived by the integral with respect to $X_{j}$
 in Eq.~(\ref{eq:pathinteg1}):
 \begin{eqnarray}
\mathcal{K}(\Upsilon _{f},\Upsilon _{i})
&=& \int\mathcal{D}\Upsilon'\mathrm{e}^{i\int_{t_{i}}^{t_{f}} \mathrm{d}t~\Tilde{\mathcal{L}} }
\label{eq:pathinteg2}
\\
\Tilde{\mathcal{L}}&=&-i\dot{\boldsymbol{\eta}}_{ t}^{\dagger} \cdot {\boldsymbol{\eta}}_{ t} 
+F\langle \mathcal{A}_{k_{t}}\rangle -\langle\hat{\xi}_{k_{t}} \rangle,\label{eq:effLag}
\end{eqnarray}
where 
$\mathcal{D}\Upsilon' =\mathcal{D}\eta^{\dagger}\mathcal{D}\eta $, and $\dot{k}=F$.

\section{Instanton method in the 1D SSH model}

Having established the path integral with the effective Lagrangian by the Bloch states, we consider the LZ problem of the lattice system.
 In a conventional way, the LZ formula of the two-state model is derived by the lowest order perturbation of the Schr$\ddot{\mathrm{o}}$dinger equation in terms of the level separation, i.e., the $k\cdot p$ approximation, by the assumption that small momenta are involved in the transition process. 
Consequently, this approach introduces the Dirac model.
However, as the gap becomes larger, the contribution beyond the linear approximation should be taken into account. 
For the lattice system, rigorous solutions are brought by the differential equation of the infinity order of the Schr$\ddot{\mathrm{o}}$dinger equation.
 When the velocity is large, the difference from the Dirac dispersion could be perturbatively treated with a finite-order differential equation, although this perturbative procedure breaks down for the small velocity, i.e., flat-band systems. 
Therefore, the nonlinearity makes solutions elusive for the conventional approach.
 Meanwhile, the path integral allows us to treat operators as c-numbers, and the higher-order effect can be analytically obtained for simple models.

 As a natural extension of the Dirac model to a 1D lattice system, we employ the SSH model \cite{Su79, Su80}.
 In the vicinity of the band edge, the SSH system is equal to the Dirac model,
 and the comparison between the lattice and Dirac systems is expected to exhibit physical effects of the periodic nature of the lattice with clarity. 
The SSH Hamiltonian is given as
 \begin{eqnarray}
H_{k}=v\sin k\alpha _{z}+\Delta\alpha_{x},
\label{eq:Ham1}
\end{eqnarray}
where the hopping integral $v$ corresponds to the velocity of the Dirac model, and we put $a=1$ for simplicity. 
The eigenvalues are given as $\pm\xi_{k}$ with $\xi_{k}=\sqrt{v^2\sin^2 k +\Delta^2}$ for the upper ($+$) or lower bands ($-$) (Fig.~\ref{fig:system1D}(a)).

For the above model, we first consider the classical equation of motion.
By using the expression for the spinor $\eta=(\cos\frac{\theta_{t}}{2}\mathrm{e}^{-i\phi_{t}}, \sin\frac{\theta_{t}}{2})^t$, the effective Lagrangian (Eq.~(\ref{eq:effLag})) of the SSH model has the form
 \begin{eqnarray}
\Tilde{\mathcal{L}} 
=\frac{1}{2}\dot{\phi}(1+\cos\theta)
+F\frac{a_{k}}{2}\sin\phi\sin\theta-\xi_{k} \cos\theta,
\label{eq:eflag}
\end{eqnarray}
where the connection is $
\mathcal{A}
=
\frac{a_{k}}{2}
\sigma_{y}
$,
 with $a_{k}= \frac{\Delta v\cos k}{\xi_{k}^2} $ and $k=Ft$. $\boldsymbol{\sigma }$ are the Pauli matrices by the basis of the eigenstates of the Hamiltonian. 
The Euler-Lagrangian equation gives the classical solutions $ \langle \boldsymbol{\sigma} \rangle =(0,0, \kappa)$ with $\kappa=\pm1$, and $\kappa$ is determined by the initial condition, i.e., the lower state $\kappa =-1$ or upper state $\kappa =1$.
Thus, the classical Lagrangian is given as 
 \begin{eqnarray}
\mathcal{L}_{\mathrm{cl}} =-\xi_{k}\kappa.
\end{eqnarray}
Since we consider the transition between the upper and lower band edges, the initial wavenumber (time) is given as $k_{i}=Ft_{i} = \frac{l\pi}{F}$ with an integer $l$. $l\pi$ comes from the valley of this tight-binding model, and since it does not affect the transition amplitude in the present problem, we choose $l=0$ in the following discussion.

We consider the action according to the instanton method by the imaginary time procedure~\cite{ColemanB}: $t\to-i\tau$.
For the classical contribution of the transition amplitude, we obtain the action by the integral of the Lagrangian along the contour on the imaginary time axis C as shown in Fig.~\ref{fig:system1D}(b).
On the contour C, the wavenumber is given by $k=iF\tau\equiv iq$, and
the magnitude of the energy is expressed as $\xi_{k}=\sqrt{\Delta^2-v^2\sinh^2 q}$.
The transition between the two adiabatic states $\pm\xi_{k}$ corresponds to the change of the branch at $\tau_{0}=\frac{1}{F}\mathrm{Arcsinh} \frac{1}{\rho} $ with $ \rho\equiv\frac{v}{\Delta}$; the band transition is brought by the change of the Riemann surface sheet but not of the spinor angle $\theta$. Then the transition amplitude $T_{\mathrm{B}}$ by the semiclassical action is given as
 \begin{eqnarray}
 T_{\mathrm{B}}&\sim&\mathrm{e}^{iS_{\mathrm{B}}},
\end{eqnarray}
with
 \begin{eqnarray}
S_{\mathrm{B}} =\int_{\mathrm{C}}\mathrm{d}t \mathcal{L}_{\mathrm{cl}}.
\label{eq:Scal}
\end{eqnarray}
The semiclassical contribution of the amplitude gives the transmission probability $P_{\mathrm{B}}=\mathrm{e}^{2iS_{\mathrm{B}}}$ for small values of $T_{\mathrm{B}}$.

The instanton action $S_{\mathrm{B}}$ (\ref{eq:Scal}) is calculated as
\begin{eqnarray}
iS_{\mathrm{B}}&=&-X_{0} \Theta_{\mathrm{B}}(\rho),\label{eq:actionB}\\
\Theta_{\mathrm{B}}( \rho)
&=&
2 \sqrt{1+ \rho ^2}\left[K\left( \frac{1}{\sqrt{1+ \rho ^2}}\right) -E\left( \frac{1}{\sqrt{1+ \rho ^2}}\right)\right],\nonumber
\label{eq:thetab}\\ 
\end{eqnarray}
where $\rho\equiv \frac{v}{\Delta}$ and $X_{0}=\frac{\Delta}{F}$ (see Appendix~\ref{ap:theta}). $K(Q)$ and $E(Q)$ are the complete elliptic integrals of the first and second kinds, respectively, with a variable $Q$ (Eq.~(\ref{eq:cei})).
The action (\ref{eq:actionB}) includes the periodic lattice effect in a non-perturbative form and is determined by two factors: $\Theta_{B}(\rho)$ and the scale of the potential barrier region $X_{0}$.
$\Theta_{B}(\rho)$ depends only by the intrinsic parameter $\rho$, and therefore, this function characterizes the magnitude of the tunneling of the material.
Figure~\ref{fig:tunnelingP}(a) shows the calculation results for $\Theta_{\mathrm{B}}$ and $\Theta_{\mathrm{D}} \equiv \frac{\pi }{2} \rho^{-1} =-\frac{iS_{\mathrm{D}}}{X_{0}}$. Then, we obtain $\Theta_{\mathrm{B}}<\Theta_{\mathrm{D}}$, and consequently the transition probability of the lattice system becomes larger than that by the LZ formula (Eq.~(\ref{eq:PLZ})), $P_{\mathrm{D}} <P_{\mathrm{B}} $. 


To see further physical difference from the Dirac model, we focus on the function $\Theta_{\mathrm{B}}(\rho) $ for the two cases $\rho\gg1$ and $\rho\ll 1$.
In those regions, $iS_{\mathrm{B}}$ is expressed as,
 \begin{eqnarray}
iS_{\mathrm{B}}(\rho)
&\sim&
-X_{0}\left[
\Theta_{\mathrm{D}}(\rho)-\frac{\pi}{16} \rho^{-3}\dots
\right],\ (\rho> 1),
\label{eq:rholarge}\\
&\sim&
-X_{0}\left[
\Theta_{\mathrm{F}}(\rho)- \rho^2 \ln\rho+\dots
\right], \ (\rho<1).
\label{eq:rhosmall}
\end{eqnarray}
$\Theta_{\mathrm{F}}(\rho)\equiv -2 + 2 \ln \frac{4\Delta}{v}$ is given by the lowest order of $\Theta_{\mathrm{B}}(\rho)$ in terms of $\rho$ and is shown in Fig.~\ref{fig:tunnelingP}(a).
For $\rho > 1$, the probability by the LZ formula is obtained by the first term, and since the second term is positive ($iS_{\mathrm{D}}=-X_{0}\Theta_{\mathrm{D}}(\rho)<0$), the transition probability definitely increases.
For $\rho \ll 1$, by the action $iS_{\mathrm{F}}\equiv -X_{0}\Theta_{\mathrm{F}}(\rho)$, Eq.~(\ref{eq:rhosmall}) gives the transition probability in another approximation form, 
$
P_{\mathrm{B}}\sim P_{\mathrm{F}} =\mathrm{e}^{2iS_{\mathrm{F}}} $ with 
 \begin{eqnarray}
P_{\mathrm{F}}= \left( \frac{ v}{4\Delta} \right)^{4X_{0}} \mathrm{e}^{4X_{0}} .\label{eq:apprx}
\end{eqnarray}
Then, $P_{\mathrm{F}}$ obviously has a different form of the function, compared with the LZ formula, and this probability is obtained when the change of the level separation $v$ (hopping or velocity) becomes smaller than the hybridization $\Delta$. By the calculation results, we note that the LZ formula and $P_{\mathrm{F}}$ can be used as the approximations for small $X_{0}$.

\section{Discussion and Experimental Observation} 
\begin{figure}[htbp]
 \begin{center}
 \includegraphics[width=60mm]{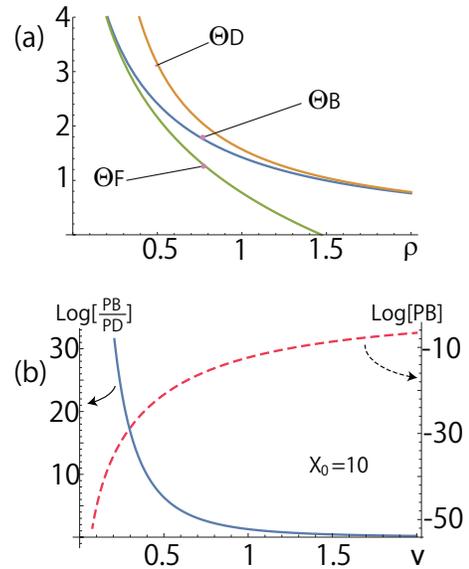}
 \caption{
(a)  The field-independent factor of the lattice model $\Theta_{\mathrm{B}}(\rho)$ in Eq.~(\ref{eq:thetab}), and its approximation forms are shown:
$\Theta_{\mathrm{D}}(\rho)\sim\Theta_{\mathrm{B}}(\rho <1) $ and $\Theta_{\mathrm{F}}(\rho)\sim \Theta_{\mathrm{B}}(\rho > 1)$.
(b) The right and left vertical axes show values of the transition probability $P_{\mathrm{B}}$ and the ratio $\frac{P_{\mathrm{B}}}{P_{\mathrm{D}}}$ in a log scale, for $X_{0}=\frac{\Delta}{F}=10$ in units of the lattice constant. $v$ is shown in units of $\Delta$.
The ratio shows $P_{\mathrm{B}}> P_{\mathrm{D}}$. 
}
\label{fig:tunnelingP}
\end{center}\end{figure}

Previously, we have shown that the tunneling probability of the lattice system becomes larger than that by the LZ formula.
Here, we first discuss the increased amount by the lattice effect.
In Fig.~\ref{fig:tunnelingP}(b), the calculation results of $P_{\mathrm{B}}$ (the right vertical axis) and the ratio of the probabilities $\frac{P_{\mathrm{B}}}{P_{\mathrm{D}}}$ (the left vertical axis) are shown for $X_{0}=10$.
By the results, we have the ratio $\frac{P_{\mathrm{B}}}{P_{\mathrm{D}}}>1$ as previously discussed.
 In addition, this ratio is very large for $\rho<1$, while $P_{\mathrm{B}}$ becomes drastically small in this region. 
For the electron in solids, we have $\rho>1$, in many cases, and by Eq.~(\ref{eq:rholarge}), the ratio of these cases becomes
\begin{eqnarray}
\frac{P_{\mathrm{B}}}{P_{\mathrm{D}}}\sim
\mathrm{e}^{\frac{\pi}{8\rho^3}X_{0} }. \label{eq:BDratio}
\end{eqnarray}
 Then, we cannot neglect the lattice effect for $\frac{\pi}{8\rho^3}X_{0} >1$.
 For the electron with the charge $-e$,  the electric field $E$ gives the force $F=- e E$, and the lattice effect appears when $E$ is smaller than $\Tilde{E}=\frac{\pi}{8}\frac{\Delta^4}{a e v^3}$.
An estimation gives $\Tilde{E}\sim 3.93\times10^3 \mathrm{Vcm}^{-1}$ for the typical parameters: $\Delta =0.1$eV, $v=1$eV, and the lattice constant $a=0.1$nm.
Therefore, the tunneling electron is affected by the lattice effect for a wide range of $E$.

We note that our calculation result differs substantially from the traditional perturbative result. 
In the conventional approach, we assume that a dispersion curve around the gap contributes mainly to the transition probability; 
namely, a lowest order Hamiltonian $H_{\mathrm{D}}$ in terms of momentum $k$ gives a principal contribution $P_{\mathrm{D}}$. 
Thus, the difference between the lattice and the lowest order dispersion is treated as a perturbative correction: $\delta P \equiv P_{\mathrm{B}} - P_{\mathrm{D}}$ with $|\frac{\delta  P }{ P_{\mathrm{D}}} |\ll1$.  
However, we obtain  the case with $\left|\frac{\delta P}{ P_{\mathrm{D}}} \right| > 1$ as shown Fig.~\ref{fig:tunnelingP}(b). 
The transition probability of the lattice is expressed as $P_{\mathrm{B}} = {\rm e}^{2iS_{\mathrm{D}}} {\rm e}^{2i\delta S}$ with $i\delta S \equiv iS_{\mathrm{B}}-iS_{\mathrm{D}}$.
Note that, although $| \frac{\delta S}{S_{\mathrm{D}}}|$ is small for $\rho>1$, $ |\delta S|$ is not always small. Therefore, ${\rm e}^{i\delta S }> 1$ can occur as discussed in the previous electron system. 
Actually, the total transition probability becomes even more than 100 times larger than that of the LZ formula as shown in Fig.~\ref{fig:tunnelingP}(b),
 and in those regions the LZ formula cannot be used.
In other wards, in the LZ tunneling problem, our result shows that the $k\cdot p$ approximation cannot be used for $\frac{\pi}{8\rho^3}X_{0} >1$.
Therefore, the transition property is not determined only by information of the band edge due to the higher-order effect.

\begin{figure}[htbp]
 \begin{center}
 \includegraphics[width=60mm]{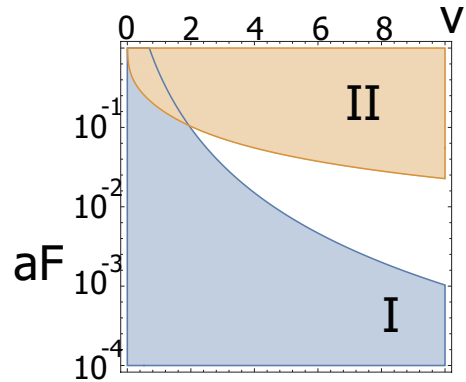}
 \caption{ Diagram for the observation of the lattice effect.
In region I$:2i\delta S-3>0 $, the strong lattice effect is obtained, and the probability becomes larger than that by the LZ formula : $\frac{P_{\mathrm{B}} }{P_{\mathrm{D}} }>\mathrm{e}^{3}\sim 20$.
In region II$:2iS_{\mathrm{B}}+\ln \bar{N}>0$ for the number of the SSH systems $\bar{N}=10^{12}$. $P_{\mathrm{B}}\bar{N}>1$ is given, and the tunneling current is assumed to be detectable.
}
\label{fig:diagram}
\end{center}\end{figure}

The lattice effect in $\Theta_{\mathrm{B}}$ can be observed directly by the tunneling.
For the observation, we consider the detection of the tunneling current with the large ratio $\frac{P_{\mathrm{B}}}{P_{\mathrm{D}}}$ 
at temperature $T$ with $k_{\mathrm{B}} T< \Delta$.
We assume an insulator stacked with the 1D SSH systems. There are $\bar{N}$ identical copies of the SSH system in the insulator along the external force, without interaction between the 1D systems.
Then, we give two criteria, I: $ 2i\delta S-3>0 $, and II: $2iS_{\mathrm{B}}+\ln \bar{N}>0$.
In region I, the lattice effect is strong, and the LZ formula cannot be used due to $\frac{P_{\mathrm{B} } }{P_{\mathrm{D}} }>\mathrm{e}^{3}\sim 20$.
Region II shows detectability of the tunneling current with $\bar{N} P_{\mathrm{B}} >1$; in reality, finite values of the tunneling probability are required for the observation. 
$\bar{N}$ is determined by the spacing $b$ between the 1 D SSH systems and the cross-section $A$ of the insulator.
 For the electron in materials, $b$ is expected to be of the order of the lattice constant.
In the calculation, we use $\bar{N}=10^{12}$, and this value is obtained for $b\sim10$ nm and  $A\sim $ 1cm$\times$1cm, for example.
The diagram of regions I and  II is shown in Fig.~\ref{fig:diagram} for the parameters of $v$ and $aF$ in units of $\Delta$. 
We find that the region satisfying the two criteria  I $ \cap $ II is limited for large $a F$ and small $v$; namely, under the criteria, the lattice effect is observable when both $X_{0}$ and $\rho$ are small.
Using $(\frac{F}{\Delta},\frac{v}{\Delta})=(0.3,0.3) \in $ I $ \cap $ II, as an example, we estimate the current as $N P_{\mathrm{B}}\sim300$.
 The approximate formula gives the close value, $\bar{N}P_{\mathrm{F}}\sim 600$.

For $ \rho>1 $, Fig.~\ref{fig:tunnelingP} indicates that it is difficult to observe the lattice effect. By Eq.~(\ref{eq:BDratio}), the condition for the large lattice effect is roughly given as $X_{0} \gg\rho^3 $.
Meanwhile, the large transition probability is obtained for $X_{0} \sim \rho $.
 Since the two conditions are contradictory to each other, the lattice effect appears with the small probability.
 Namely, the strong lattice effect appears with small values of the transition probability, and the direct observation requires high sensitivity detectors or insulators with a large number of the 1D LZ systems.
 For $ \rho<1 $, we will observe the lattice effect in this region, although the transition probability drastically decreases as shown in Fig.~\ref{fig:tunnelingP} (b).
For the electron in solids, the condition $ \rho <1$ is expected in some systems with the strong correlation~\cite{MottB}, for example. In addition, recently, the flat-band dispersion has been studied in carbon nanotube superlattices \cite{Chiko05, Chiko08}.
In these systems, $\rho$ can be tuned\cite{RibeiroRM}, and by changing the chemical potential or structure of superlattices, small values of $X_{0}$ and $\rho$ are expected.
In the SSH optical lattices systems with an applied external field \cite{Atala13},
the physical parameters are expected to be tuned, and the strong $F$ will be obtained by the large gradient of the magnetic field.
Then, the experimental observation will be done by using these systems.

In summary, we theoretically investigate the LZ problem of the Bloch states, and the enhancement of the tunneling probability is shown by the periodic nature of the lattice. 
In many cases, although particles in lattice show complex band structures, the system can be reduced to the present 1D SSH model for the particle with the energy close to the gap.
Therefore, in reality, the tunneling probability of lattice systems is expected to be underestimated by the LZ formula. 
In this study, the lattice effect emerges by the spatial gradient of the potential
 on bulk periodic systems; namely the periodicity is broken by the force.
 Recently, transport phenomena of magnons by the gradient of the magnetic field has been studied in the presence of the gap~\cite{Takahashi16}, and for these systems, the lattice effect is also expected.
 
 At the same time, the periodic effect can be regarded as a nonlinear effect.
For standard LZ problems, an explicit time-dependent driving in the Hamiltonian is 
considered~\cite{Zener32,Wernsdorfer99, Damski05}.
In these systems, the change of the force can be nonlinear by an external control, and the magnitude of the nonlinear effect on the transition amplitude  cannot be neglected when its contribution to the action is larger than unity, even though this value is much smaller than the Dirac action.  
Recently, for the many-body effect of the LZ problems, the Coulomb potential has been studied~\cite{Ostrovsky03,Sinitsyn13},
 and this interaction gives different behavior of the tunneling.
 The lattice effect will give a correction for the Coulomb problem, and this problem is important for electron systems, since the correlation becomes strong in some materials.
General systems inevitably involve nonlinear or perturbative effect, e.g., interaction between particles, disorders, and high-energy effects as shown in this study,  and similar effects are expected in ubiquitous quantum tunneling, besides periodic systems.

\begin{acknowledgments}
We thank R. Shindou for useful discussions.
\end{acknowledgments}
\appendix
\section{Calculation for $\Theta_{\mathrm{B}}$}\label{ap:theta}
Here, we give a derivation $\Theta_{\mathrm{B}}=-\frac{iS_{\mathrm{B}}}{X_{0}}$ in Eq.~(\ref{eq:thetab}) in detail. 
The instanton action is given as
  \begin{eqnarray}
iS_{\mathrm{B}} &=&2\frac{1}{F} \int_{0}^{-q_{0}}\mathrm{d}q\sqrt{\Delta^2-v^2\sinh^2q }.
\end{eqnarray}
with $q_{0}=\mathrm{Arcsinh}(\rho^{-1})$.
By putting $x=\sqrt{1-\left(\frac{v}{\Delta}\right)^2\sinh^2q }>0$, the action takes the form
  \begin{eqnarray}
iS_{\mathrm{B}} &=&2\frac{\Delta }{F} \int_{0}^{-q_{0}}\mathrm{d}q  x
=2\frac{\Delta }{F} [ q x]^{q=-q_{0}}_{q=0}- 2\frac{\Delta }{a F} \int_{0}^{-q_{0}}\frac{\mathrm{d}x}{\mathrm{d}q} q\nonumber\\ 
&=&-2\frac{\Delta }{F} \int_{1 }^{0}\mathrm{d}x q(x)
\nonumber\\
&=& -2\frac{\Delta }{F} \int^{1 }_{0}\mathrm{d}x \mathrm{Arcsinh}\left[\left(\frac{\Delta}{v}\right)\sqrt{1-x^2}\right],
\end{eqnarray}
where we chose $q(x)= -\mathrm{Arcsinh}\left[\left(\frac{\Delta}{v}\right)\sqrt{1-x^2}\right]$ for $q<0$.
The above equation reflects the energy conservation $\xi \kappa+ F X=$const., and for the Dirac model the similar calculation has been done by using the WKB method~\cite{ZimanB72}.
By transformation of the variable
$\cos \lambda=\sqrt{1-x^2}$, $\Theta_{\mathrm{B}}$ (\ref{eq:thetab}) is calculated as
 \begin{eqnarray}
\Theta_{\mathrm{B}}
&=&2\int_{0}^{1}\mathrm{d}x\mathrm{Arcsinh}(\rho^{-1} \sqrt{1-x^2})
\nonumber\\
&=&2\int_{0}^{\frac{\pi}{2}}\mathrm{d}\lambda \cos\lambda \mathrm{Arcsinh}(\rho^{-1} \cos\lambda)\nonumber\\
&=&2\int_{0}^{\frac{\pi}{2}}\mathrm{d}\lambda \cos\lambda \ln(\rho^{-1} \cos\lambda +\sqrt{1+\rho^{-2} \cos^2\lambda }),\nonumber
\end{eqnarray}
where we used $\mathrm{Arcsinh}(x)=\ln(x+\sqrt{1+x^2})$, and
 \begin{eqnarray}
\Theta_{\mathrm{B}} &=&
=-2\rho^{-1}  \int_{0}^{\frac{\pi}{2}}\mathrm{d}\lambda\sin^2 \lambda 
\frac{1 +\frac{\rho^{-1} \cos\lambda}{\sqrt{1+\rho^{-2} \cos^2\lambda} }}
{\rho^{-1} \cos\lambda +\sqrt{1+\rho^{-2} \cos^2\lambda }}\nonumber\\
&=&
2\rho^{-1}  \int_{0}^{\frac{\pi}{2}}\mathrm{d}\lambda
\frac{\sin^2 \lambda } {\sqrt{1+\rho^{-2} \cos^2\lambda} }
\nonumber\\
&=&\frac{2\sqrt{1+\rho^{-2} }}{\rho^{-1} }\left[K\left( \sqrt{\frac{1 }{1+\rho^{2} }}\right)  -E\left( \sqrt{\frac{1 }{1+\rho^{2} }}\right) \right],\nonumber\\
\end{eqnarray}
where $K(Q)$ and $E(Q)$ are the complete elliptic integrals of the first and second kinds:
 \begin{eqnarray}
K(Q)&\equiv&\int_{0}^{\frac{\pi}{2}}\mathrm{d}\theta\frac{1}{\sqrt{1-Q^2\sin^2\theta}}\nonumber\\
E(Q)&\equiv&\int_{0}^{\frac{\pi}{2}}\mathrm{d}\theta\sqrt{1-Q^2\sin^2\theta} .
\label{eq:cei}
 \end{eqnarray}


\begin{thebibliography}{99}
\bibitem{Landau32}
L. D. Landau, Phys. Z. Sowjetunion, \textbf{2} , 46 (1932).
\bibitem{LandauB}
L. D. Landau and E. M. Lifshitz, in Quantum Mechanics: Nonrelativistic Theory, (Pergamon, New York, 1977)
\bibitem{Zener32}
C. Zener, Proc. R. Soc. London, Ser. A \textbf{137} , 696 (1932)
\bibitem{Spreeuw90}
R. J. C. Spreeuw, N. J. van Druten, M. W. Beijersbergen, E. R. Eliel, and J. P. Woerdman, {Phys. Rev. Lett.} { \bf 65}, 2642 (1990) .
\bibitem{Bouwmeester95}
D. Bouwmeester, N. H. Dekker, F. E. v. Dorsselaer, C. A. Schrama, P. M. Visser, and J. P. Woerdman, Phys. Rev. A  {\bf 51}, 646 (1995) .
 \bibitem{NiuQ00}
 B. Wu and Q. Niu, Phys. Rev. A \textbf{61}, 023402(2000)

 \bibitem{Liu02}
 J. Liu, et al.  Phys. Rev. A \textbf{66}, 023404 (2002).

\bibitem{Chen11}
  Y.-A. Chen et al., et al., Nature Phys.\textbf{7},61 (2011).

\bibitem{ChildB}
Child, M. S. {\it Molecular Collision Theory} Academic Press, London (1974).
\bibitem{NikitinB} Nikitin, E. E. \& Umanski, S. Y. {\it Theory of Slow Atomic Collisions} (Springer, 1984).


\bibitem{Asamitsu97}
A. Asamitsu, Y. Tomioka, H. Kuwahara, and Y. Tokura, Nature( London )\textbf{388}, 50 (1997).
\bibitem{Miyano97}
K. Miyano, T. Tanaka, Y. Tomioka, and Y. Tokura, Phys. Rev. Lett. \textbf{78}, 4257 (1997).
\bibitem{Yamanouchi99}
S. Yamanouchi, Y. Taguchi, and Y. Tokura, Phys. Rev. Lett. \textbf{83}, 5555 (1999).

\bibitem{OkaT05}
T. Oka and H. Aoki, Phys. Rev. Lett. \textbf{95}, 137601(2005).
\bibitem{Sugimoto08}
N. Sugimoto, S. Onoda, N. Nagaosa, Phys. Rev. B \textbf{78}, 155104 (2008 ).
\bibitem{Taguchi00}
Y. Taguchi, T. Matsumoto, and Y. Tokura Phys. Rev. B \textbf{62}, 7015 (2000)

\bibitem{ZimanB72}
 Ziman, J. M. {\it Principles of the Theory of Solids} 2nd edn (Cambridge Univ. Press, 1972).
 \bibitem{SakuraiB}
Sakurai, J. J. {\it Modern Quantum Mechanics}. (Addison-Wesley, 1993). .

\bibitem{FeynmanB}
Feynman, R. P. \& Hibbs, A. R.  {\it Quantum Mechanics and Path Integrals} (McGraw-Hill, 1965).



\bibitem{ColemanB}
Coleman,  S. {\it Aspects of symmetry} (Cambrides, 1985). 
\bibitem{Su79} 
 W. P. Su, J. R. Schrieffer, and A. J. Heeger, Phys. Rev. Lett. \textbf{42}, 1698 (1979).

 \bibitem{Su80}
 W. P. Su, J. R. Schrieffer, and A. J. Heeger, Phys. Rev. B \textbf{22}, 2099 (1980).

\bibitem{Culcer05}
 D. Culcer, Y. Yao, and Q. Niu Phys. Rev. B \textbf{72}, 085110 (2005).
\bibitem{Xiao10}
D. Xiao, M. Chang, and Q. Niu, { Rev. Mod. Phys.} \textbf{82}, 1959 (2010).

 



\bibitem{MottB}
Mott, N. F. {\it Metal-Insulator Transitions} (Taylor \& Francis, London, 1990).




\bibitem{Chiko05}
W. Jask\'olski and L. Chico,  Phys. Rev. B, \textbf{71},155405(2005).
\bibitem{Chiko08}
A. Ayuela, L. Chico, and W. Jask\'olski, Phys. Rev. B.\textbf{77},085435(2008).

\bibitem{RibeiroRM}
R. M. Ribeiro, V. M. Pereira, N. M. R. Peres, P. R. Briddon, and A. H. C. Neto, New J. Phys. \textbf{11}, 115002 (2009).

\bibitem{Atala13}
M. Atala, M. Aidelsburger, J. T. Barreiro, D. Abanin, T. Kitagawa, E. Demler, and I. Bloch, Nat. Phys. \textbf{9}, 795 (2013).
\bibitem{Takahashi16} R. Takahashi and N. Nagaosa, Phys. Rev. Lett. \textbf{117}, 217205 (2016).
\bibitem{Wernsdorfer99} W. Wernsdorfer and R. Sessoli, Science \textbf{284}, 133 (1999).
\bibitem{Damski05} B. Damski, Phys. Rev. Lett. \textbf{95}, 035701(2005).


\bibitem{Ostrovsky03} V. N. Ostrovsky, Phys. Rev. A \textbf{68}, 012710 (2003).
\bibitem{Sinitsyn13}N. A. Sinitsyn, Phys. Rev. Lett. \textbf{110}, 150603 (2013).

\end{thebibliography}



\end{document}